# Beyond-CMOS Device Benchmarking for Boolean and Non-Boolean Logic Applications

Chenyun Pan, *Member*, *IEEE*, and Azad Naeemi, *Senior Member*, *IEEE*

*Abstract*—The latest results of benchmarking research are presented for a variety of beyond-CMOS charge- and spin-based devices. In addition to improving the device-level models, several new device proposals and a few majorly modified devices are investigated. Deep pipelining circuits are employed to boost the throughput of low-power devices. Furthermore, the benchmarking methodology is extended to interconnect-centric analyses and non-Boolean logic applications. In contrast to Boolean circuits, non-Boolean circuits based on the cellular neural network demonstrate that spintronic devices can potentially outperform conventional CMOS devices.

## I. INTRODUCTION

FACED with the challenges and limitations of CMOS scaling, there is a global search for beyond-CMOS device technologies that are capable of augmenting or even replacing conventional Si CMOS technology and sustaining Moore's Law [1-4]. There is an increasing need for a uniform benchmarking methodology to capture and evaluate the latest research and development for various beyond-CMOS proposals. Such research is critical in identifying the key limiting factors for promising devices and in guiding future research directions through modification or even reinvention of proposed devices.

Beyond-CMOS Benchmarking (BCB) efforts have continued for several years with three major releases. The first one originating in 2010, BCB 1.0, was led by K. Bernstein [2] and used unmodified device inputs from various NRI groups. It was followed by two sequential uniform benchmarking works led by D. Nikonov and I. Young [3, 4], BCB 2.0 and 3.0. They treated a broader range of devices and circuits using a consistent, transparent, and physics-based methodology [4].

In this paper, we have added two recently proposed voltage-controlled spintronic devices, magnetoelectric magnetic tunneling junction (MEMTJ) and composite–input magnetoelectric–based logic technology (CoMET). More elaborate device-level modeling approaches are applied to many spintronic devices. In addition, major modifications for several spintronic devices have been proposed and evaluated. Major updates have also been applied for charge-based FETs to reflect the latest research developments in the past two years, i.e. tunneling FET (TFET), ferroelectric-based FET, graphene pn junction device, 2D material based devices, and negative differential resistance (NDR) devices.

At the circuit level, the arithmetic logic unit (ALU) circuit is adopted from the previous Boolean logic benchmarking [4]. We include the deep pipelining analysis to take advantage of the inherent memory feature of some of the low-power devices and the supply clocking that has to be used to eliminate standby power dissipation in current-driven devices. Since this approach is somewhat similar to dynamic logic, CMOS implementation of dynamic logic has been added to the reference benchmarking data points. To account for the fact that interconnects pose major limitations on the state-of-the-art VLSI systems [5, 6], interconnect-centric performance benchmarking is also included in this paper. It covers multiple key interconnect metrics, such as the optimal delay and energy of a long interconnect with repeaters and the span-of-control. We then explore the benefits and limitations of emerging charge- and spin-based technologies from the perspective of interconnect design. Following BCB 3.0, we choose a simple analytical approach to capture the key advantages, challenges, and limitations of various emerging technologies.

Previous benchmarking results for a 32-bit ALU have shown that only a few devices can potentially outperform CMOS in terms of energy-delay product (EDP) [4]. Most devices are worse in terms of delay and energy per ALU operation, especially for spintronic devices, where orders of magnitude larger EDPs are projected. Therefore, it is crucial to search for non-traditional circuits where beyond-CMOS devices can realize their full potential. During the past few years, research has shown that alternative computing paradigms, such as non-Boolean circuits and systems, are potentially capable of taking advantage of the unique physical properties of novel devices [7, 8]. In this paper, we choose the cellular neural network (CNN) as the benchmarking circuit because: 1) it performs many tasks in the areas of sound, image, and video processing quite efficiently [9, 10]; 2) it has a well-established theory [11]; and 3) it can be implemented by a wide range of emerging technologies for both charge- and spin-based devices [9, 10, 12]. Furthermore, a recent work has shown that cellular neural network can be efficiently used to create a convolutional neural network that is widely used in deep-learning applications [13]. In this work, we investigate three types of CNN

Manuscript submitted October 25, 2017. This work was supported by the Semiconductor Research Corporation (SRC) NRI Theme 2624.001.

C. Pan, and A. Naeemi are with the School of Electrical and Computer Engineering, Georgia Institute of Technology, Atlanta, GA 30332 USA (e-mail: chenyun.pan@gatech.edu).

implementations, i.e. analog, digital, and spintronic circuits, covering a broad range of charge- and spin-based beyond-CMOS devices.

The rest of the paper is organized as follows. Section II introduces several new devices and latest device-level models to be covered in this new release of benchmarking. Section III shows the circuit-level benchmarking methodology for both Boolean and non-Boolean computing. Benchmarking results and discussions are presented in Section IV. Finally, conclusions are made in Section V.

## II. Device-Level Models

### A. Charge-Based Devices

*1) Tunneling FETs*

The intrinsic capacitance and ON current of several TFETs have been modified significantly for two-dimensional heterojunction interlayer TFET (ThinTFET), Gallium Nitride TFET (GaNTFET), and transition metal dichalcogenide TFET (TMDTFET) [14-16] (Figure 1 Supplementary material). These modifications are based on atomistic simulations performed at the Low Energy Systems Technology (LEAST) center, a research center sponsored by SRC and DARPA through STARnet [17]. Substantial performance improvements are observed for the following reasons. The wave function coupling between the two 2D materials in ThinTFET has been adjusted to the best-known value. An updated simulation of the charge in the device has changed the gate capacitance. The 2015 NEMO simulation [18] of the inline GaNTFET at 0.4 V has resulted in a much larger saturation current than the comparable TCAD simulation. In addition to improvements in the atomistic simulations using NEMO, there have been changes in materials, structures, stress, and doping of TMDTFET. The simulations have been performed for 15 device options, and the best ones are selected in this paper [17]. The baseline TFET devices, i.e. the homogeneous and heterogeneous TFETs, are taken from BCB 3.0 [4].

*2) Ferroelectric-Based FETs*

Three ferroelectric-based FETs from BCB 3.0 are included in this work: negative capacitance FET (NCFET), metal-insulator transition FET (MITFET), and ferroelectric FET (FEFET) [4, 19, 20]. The updated IV characteristics for the NCFET is obtained from the LEAST center [21, 22]. Following [4], a partial polarization intrinsic switching time of 10 ps is added on top of the intrinsic switching delay of NCFET. One change made in this benchmarking release is that only one ferroelectric switching delay is added on different logic function, such as an inverter and NAND gate, assuming the ferroelectrics on the gates of NMOS and PMOS transistors are switched in parallel.

*3) Other Charge-Based Devices*

There is a major update on the modeling approach of the graphene pn junction (GPNJ) devices. Instead of the analytical angular dependent transmission probability model used in previous benchmarking, a more realistic model based on ray-tracing approach is employed [23]. The results are more accurate and consistent with rigorous NEGF simulations. The ON-OFF ratio of the device predicted by the new model degrades by 10× even at a large device width because of multiple reflections of electron beams at graphene edges and junctions. Two configurations of GPNJ devices are investigated with different input backgates.

Two negative differential resistance (NDR) devices, bilayer pseudospin FET (BisFET) and interlayer tunneling FET (ITFET), from BCB 3.0 are included in this work [4, 24]. After voltage signals reach the input of logic gate, complementary supply voltages are applied to perform the logic computation and lock the output according to the inputs. The supply voltage needs to be held in order to lock the output voltage. Meanwhile, the logic device consumes static power until the supply voltage returns to zero. The 2D material-based van der Waals FET (vdWFET) device is updated based on a new channel material, black phosphorus, which has a large field-effect mobility and highly anisotropic bandstructure [25].

### B. Spintronic Devices

Another category of devices is the spintronic devices. These devices are promising candidates to complement conventional CMOS devices as they provide new features, such as non-volatility and low-voltage operation [26]. One set of the spintronic devices are current-driven, and some of the well-studied device concepts in this category include all-spin logic (ASL) [27], charge-coupled spin logic (CSL) [28], and domain wall logic (mLogic) [29]. Another set of spintronic devices is based on voltage-controlled switching of magnets. These devices can potentially improve energy efficiency because they do not need a large current and avoid the energy associated with the Joule heating and the leakage. The representative devices included in this work are the MEMTJ device, the spin wave device (SWD), and CoMET.

MEMTJ and CoMET are two new devices that are added to the benchmarking. Furthermore, several modified technology options for MEMTJ and CSL devices with more advanced device materials and structures have been evaluated. More accurate modeling approaches have also been used for the existing devices, such as ASL, CSL, mLogic, and SWD. The updated modeling approach for each device is described as follows.

*1) All-Spin Logic (ASL)*

The original ASL device was proposed in [27]. Compared to the model used in BCB 3.0, the new model takes into account spin relaxation during the spin diffusion along the metallic channel. The spin polarized current density received at the output magnet is [30]

$$J_s = \frac{\beta J_c}{\frac{\sinh(l_c/l_{sf})\cosh(l_g/l_{sf})}{\sinh(l_g/l_{sf})} + \cosh(l_c/l_{sf})}, \quad (1)$$

where $J_c$ is the charge current density, $l_c$ is the channel length, $l_g$ is the length of the ground path, $\beta$ is the spin injection coefficient, and $l_{sf}$ is the spin diffusion length. The spin diffusion length in copper has been obtained using compact physical models that account for surface and ground boundary scatterings in nanoscale wires [31]. The spin injection coefficient, critical switching current, and magnet switching delay are adopted from previous work [4]. In addition to the in-plane magnet, perpendicular magnetic anisotropy (PMA) magnets are also investigated in ASL devices. PMA magnets require much lower critical switching currents, allowing for more energy efficient computing [32]. One novel PMA magnetic material, Heusler alloy, is also added into the benchmarking plot. It has the advantage of a large anisotropy value of $2.6 \times 10^6$ J/m$^3$, allowing a small magnet size without sacrificing thermal stability and enabling a fast switching time [33].

*2) Charge-Spin Logic (CSL)*

CSL was originally proposed in [28], as shown in Fig. 1 (a). The magnetic orientation of the bottom free magnet is controlled by the spin orbital torque generated by passing a charge current through a heavy metal, namely spin Hall effect (SHE). The magnetic orientation of the top magnet is controlled by the bottom magnet via dipole coupling. The two magnets are electrically isolated which ensures input/output isolation. The magnetization of the top magnet determines the polarity of the charge current transmitted to the next stage via the tunneling magnetoresistance (TMR) effect. Several updates have been made for the CSL device: the magnets were made 3× bigger to fit two MTJs; the magnet parameters were updated to ensure perfect dipole coupling [34]; and the output resistance network is included to estimate the driving current to the next stage, as shown in the supplementary material.

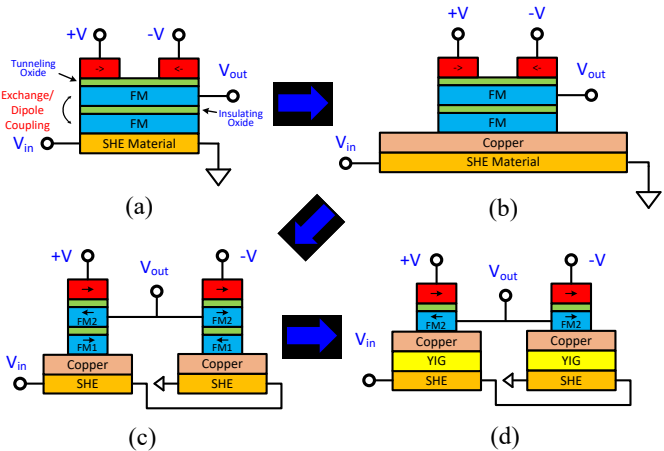

Fig. 1. Schematics of four CSL devices (a) with originally proposed device, (b) that use copper collector to boost spin current, (c) that separate the pull-up and pull-down network, and (d) that add a YIG layer between the copper collector and SHE material.

To further improve the spin current injected into the bottom magnet, a copper layer has been proposed to collect spins from a large surface area and funnel them to the magnet via spin diffusion [35], as shown in Fig. 1 (b). For the benchmarking results shown in Section IV, an enhancement factor of 2 is considered to show the potential improvement by adding the copper collector (supplementary material). This factor has been calculated by accounting for the fact that the electrical current in the heavy metal gets partly shunted by the copper layer. Using a thinner copper layer decreases the shunted electrical current; however, the spin diffusion length in the copper layer decreases due to size effects [31]. Hence, there is an optimal Cu thickness that maximizes the spin transfer torque for a given electrical current.

For the device structure shown in Fig. 1 (a) and (b), one major fabrication challenge is to create two fixed magnets side by side whose magnetizations point in opposite directions. To address this challenge, we propose and investigate a new CSL structure by breaking the device into two complementary pull-up and pull-down networks, as shown in Fig. 1 (c). In this complementary device, all fixed magnets point in the same direction. The last advancement we investigated is adding an Yttrium iron garnet (YIG) layer between the SHE material and the copper collector, as shown in Fig. 1 (d). This creates an insulating layer that electrically isolates the input and the output, eliminating the need for dipole coupling between two free magnets [36]. In addition, the YIG layer prevents the parasitic current path through the copper collector and further improves the spin injection by an extra 50% [37]. The comparison among four different CSL devices with advanced materials and structures will be illustrated in Section IV.

*3) Magnetic Domain Wall Magnetic Logic (mLogic)*

The magnetic domain wall based logic device, mLogic, is included in the updated benchmarking work. It is considered as the same device as the STT-DW device in BCB 3.0 but with an updated complementary logic implementation and numerical simulation for the domain wall velocity. Unlike other spintronic devices relying on majority-gate logic, mLogic devices perform computation with complementary logic circuits that are similar to CMOS circuits. The output voltage depends on the pull-up and pull-down resistance networks that are set according to the input current generated by the previous stage. The modeling approach, such as the domain wall speed versus the input current density, follows the previous work [29].

*4) Magnetoelectric Magnetic Tunneling Junction (MEMTJ) Device*

The proposed stand-alone voltage-controlled MEMTJ device is shown in Fig. 2 (a). The original MEMTJ logic concept was originated in [38, 39]. The basic building block consists of a magnetoelectric antiferromagnetic (AFM) layer stacked with an MTJ. Chromia ($Cr_2O_3$) provides an exciting opportunity in this regard. The boundary magnetization of $Cr_2O_3$ can be isothermally controlled via an applied electric field and the generated voltage-controlled perpendicular exchange bias can be used to switch an adjacent ferromagnetic layer [40-43]. The magnetization of the free magnet determines the output MTJ resistance. Using a MOSFET at the output, this MTJ resistance is converted back to the voltage and drives the next stage. Building upon this design, we propose a stand-alone voltage-controlled magnetoelectric device to address the following challenges and limitations. First, each device needs multiple dedicated MOSFETs to drive the next stage. Second, a preset and clocking scheme is required to perform logic functions since the output voltage is only positive. Third, devices are very sensitive to the insulator thickness variability because the voltage division between the FET and the MTJ determines the output voltage. Any variation in the insulator thickness changes the MTJ resistance exponentially, and consequently shifts the output voltage significantly.

The proposed MEMTJ is similar to the CSL device in which the current controlled write element (SHE) has been replaced with a voltage-controlled magnetoelectric element. Like CSL, it satisfies all five essential requirements of general logic applications: nonlinearity, gain, concatenability, feedback prevention, and a complete set of Boolean operations based on the majority gate and inverter. However, the magnetoelectric effect is far more energy efficient compared to spin transfer torque. The proposed device can directly drive the next stage and eliminate the need for any auxiliary FETs. Two matched MTJs are built at the output stage so that the output voltage is not sensitive to the MTJ insulator thickness variability. This is because the pull-up and pull-down resistances change proportionally as the MTJ insulator thickness changes. A major challenge for this device is to ensure perfect coupling between the write and read magnets via dipolar coupling. MEMTJ uses PMA magnets and the design space for perfect coupling of PMA magnets via dipole coupling is quite narrow [44].

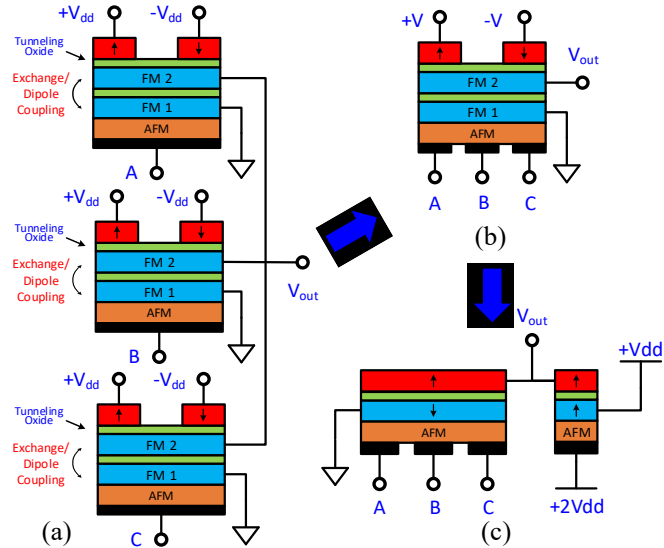

Fig. 2. Schematics of three MEMTJ devices. (a) standard MEMTJ, (b) compact MEMTJ device with the assumption of a single-domain magnet, and (c) preset-based MEMTJ device without the dipolar coupling.

One can connect the outputs of three MEMET to create a majority gate. However, a more compact MEMTJ device can be created assuming a single-domain magnet similar to the previous proposal [41, 42] (Fig. 2 (b)). In this case, the AFM is gated with three inputs and the magnetization of the free bottom magnet depends on the majority of the inputs. This compact device option also improves the voltage swing of the output because the pull-up and pull-down units only have one MTJ. In addition, the input capacitance is smaller compared to the case where the outputs of three inverters (Fig. 2 (a)) are connected in parallel.

To address the challenge with the dipole coupling and also to eliminate the need for fixed magnets pointing in opposite directions, a third MEMTJ device is proposed in Fig. 2 (c). The detailed operation and configuration during the preset and computation period are described in the supplementary material.

*5) Spin Wave Device (SWD)*

In SWD, a voltage is applied across a piezoelectric material to create strain and change the magnetization of the magnet through a magnetostrictive effect. This creates a spin wave that propagates through the magnetic channel toward the output magnet. The output magnet is preset at the meta-stable condition until the spin wave arrives. After that, a phase-dependent deterministic switching is realized by modifying the energy landscape and shifting the location of the saddle point [45]. A clocking scheme enables the correct detection and transmission of spin wave signals and guarantees non-reciprocity [46]. The majority of the delay is associated with the magnet switching from the metastable state to the steady state. The intrinsic switching energy of the ME cell dominates the overall energy dissipation. The energy associated with the clocking is small because one clocking transistor can drive hundreds of ME cells in the same stage. The detailed modeling approaches to calculate the energy and delay per operation is described in [47].

*6) Composite–Input Magnetoelectric–Based Logic Technology (CoMET)*

CoMET is another new voltage-controlled device that has been added into the benchmarking. It enables low-voltage-induced domain wall nucleation based on the magnetoelectric effect. The fast domain propagation is realized by passing a charge current through a heavy metal laid underneath the PMA magnetic channel (spin Hall effect). The delay and energy dissipation are dominated by the domain wall nucleation/propagation and the switching energy of CMOS transistors, respectively. The modeling approaches to calculate the delay per operation is adopted from the numerical simulation in [48] with updated energy dissipation calculation for the dynamic switching energy of the transistors, leakage energy of CMOS inverters, and Joule heating energy.

*C. Archived Devices*

Some of the devices from BCB 3.0 are archived due to their uncompetitive performance as well as the lack of activity in various research centers. These devices include graphene nanoribbon FET (GNR TFET), SpinFET, piezoelectric FET (PiezoFET), excitonic FET (ExFET), spin majority gate (SMG), nanomagnetic logic (NML), and spin torque oscillator (STOlogic).

III. BENCHMARKING METHODOLOGY

*A. Boolean Logic Benchmarking*

The 32-bit ALU from BCB 3.0 [4] is adopted for the circuit-level benchmarking of the Boolean circuits. One major update has been made for the clocking scheme of NDR devices (i.e. BisFET and ITFET). Complementary logic gates based on NDR devices take their input at the rising edge of the supply voltage and their output will not change until the supply voltage falls to zero again. A multiphase clocking scheme has been proposed to perform logic computation and propagation in a multistage Boolean circuit, such as the ripple carry adder [49]. To ensure all SUM bits are available at the end of the 32-bit addition, BCB 3.0 assumes that all NDR devices are constantly clocked until all 32 bits are computed. This constant clocking means that each bit of sum is calculated 32 times, and it is used only once at the end. Alternatively, in this paper we disable the clocking after each logic gate finishes the computation, and only hold the clock on for the last logic gates (XOR) of each full adder to lock the SUM bit. This reduces the dynamic power dissipation by 32× but increases the leakage power of the last logic gates. The implications of this trade-off will be discussed in Section IV.

To better utilize the power density budget and quantify the benefit of the deep-pipelining, the standard N-P domino logic is implemented to boost the throughput of low-power FETs, as shown in Fig. 3 [50]. The delay is estimated based on the worst-case input combinations, and the energy is estimated based on the switching probability of inputs as well as internal nodes, such as the $\overline{Carry}$ and $\overline{Sum}$.

For the NDR devices, BisFET and ITFET, as well as the spintronic devices, the supply clocking that needs to be used to ensure functionality or to eliminate standby power enables a similar pipelining where the logic depth becomes equal to one. All these devices are intrinsic memory elements. Magnetic devices are non-volatile, and NDR devices latch the input signal only at the rising edge of the supply voltage (clock). The value of the input has no effect on the output so long as the supply voltage remains high.

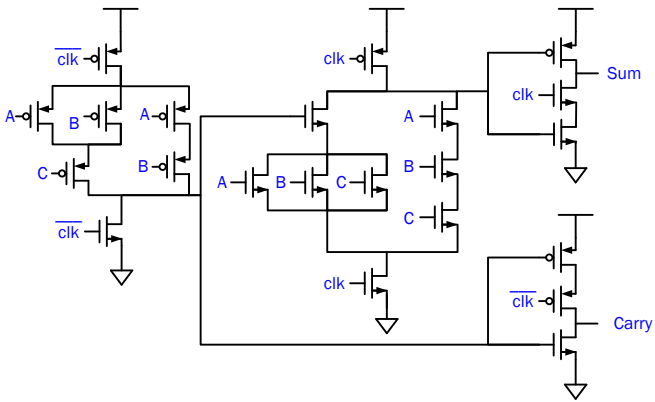

Fig. 3. Circuit diagram of a full-bit adder using the standard N-P domino logic [50].

*B. Interconnect Centric Analysis*

Interconnects impose a major limitation on the state-of-the-art integrated circuits. Previous studies have shown that interconnects account for more than half of dynamic power dissipation and critical path delay [5]. More than 50% of the logic cells on a chip may be used as repeaters for long interconnects [6]. As wire dimensions scale, the size effect significantly increases wire resistivity below sub-20nm nodes [51]. Therefore, it is crucial to evaluate the implications of various novel device proposals. The energy and delay of an interconnect with the optimal repeater insertion are derived as [52]

$$t_{wire} = 1.4\sqrt{R_0 C_0 r_w c_w} \cdot l + 2\sqrt{(0.7 R_0 C_0 + t_p) 0.4 r_w c_w} \cdot l \quad (2)$$

$$E_{wire} = \frac{1}{2} c_w l \left(1 + \sqrt{\frac{0.4 R_0 C_0}{0.7 R_0 C_0 + t_P}}\right) V_{dd}^2 \quad (3)$$

respectively, where $r_w$ and $c_w$ are the resistance and capacitance per unit length of the interconnect, $l$ is the length of the interconnect, $t_p$ is the extra polarization switching time of ferroelectric devices, $R_0$ and $C_0$ are the output resistance and input capacitance of a minimum-sized repeater, respectively, and $V_{dd}$ is the supply voltage.

The span of control, originally proposed in [53], addresses the communication among logic switches by measuring the number of accessible devices within one clock cycle. This metric is a rough indicator of the circuit block size for each device technology beyond which interconnects become a major limitation. In this work, the clock period is assumed to be 300 times the intrinsic device delay. Here, only static logic circuits are considered. Results based on the latest device-level models in Section II will be discussed in Section IV.

*C. Neuromorphic Benchmarking Circuits*

Despite the large research efforts in the Boolean logic domain, few device concepts have better or comparable performance compared to conventional CMOS technology [4]. Recent studies have shown that non-Boolean logic can better utilize emerging technologies, such as spintronics, and achieve a better computing energy efficiency [7, 8].

Some non-Boolean computing architectures, such as the CNN [11], are promising candidates to provide higher energy efficiencies due to their massively parallel processing capability. In this work, we follow the same methodology presented in our previous work to benchmark a variety of charge- and spin-based devices [54]. For charge-based devices, both analog and digital implementations are investigated. Results are simulated based on the updated device-level characteristics, i.e. the bias current, sub-threshold slope, and supply voltage. For the spintronic CNN, devices with new materials and structures are included, including ASL devices with Heusler alloy magnets and CSL devices with copper collector and YIG. The associative memory application is investigated for three types of CNN implementations using 4-bit weight synapses to achieve 90% recall accuracy for a given input noise of 10%.

IV. RESULTS AND DISCUSSIONS

In this section, the benchmarking results are demonstrated based on the device-, interconnect-, and circuit-level models described in Sections II and III.

## A. 32-bit Adder

### 1) Energy vs. Delay

Results of a 32-bit ALU are shown in Fig. 4 for a variety of charge- and spin-based devices. In general, spintronic devices consume more energy and delay per operation due to the large switching delay of nanomagnets as well as the large Joule heating of the current-driven devices.

Compared to the previous benchmarking work, the data points for several TFETs have moved considerably toward the preferred corner because of more accurate modeling approaches and improved material and structures. Compared to other TFET devices, ThinTFET provides the best performance thanks to its steep subthreshold slope and large ON current at a small supply voltage. The ultra-thin channel structure leads to a strong gate control over the channel. Layered 2D crystals provides a sharp turn on of density of states at the band edges and have no surface dangling bonds. This potentially enables a low interfacial density of state, which are highly desired for achieving a steep subthreshold slope [55]. GPNJ devices consume a much larger energy per operation because the new modeling approach based on ray-tracing and NEGF has demonstrated that a larger device size is required to achieve a reasonable ON-OFF ratio, leading to large dynamic switching energy. For BisFET and ITFET, a new clocking scheme is applied, as described in Section III A. Since the supply clocking of the logic gate is disabled once the computation completes, the dynamic energy is reduced significantly. However, with a low ON-OFF ratio of 10, BisFET and ITFET suffer from large leakage energy, which contributes to the majority of the energy dissipation.

corner because of the more realistic modeling approach described in Section II. However, many recent advancements in both device structures and materials improve the switching delay and reduce the critical switching current requirements, leading to a continuous performance improvement towards the preferred corner. One can observe that voltage-controlled spintronic devices, such as MEMTJ, SWD, and CoMET, have a great advantage in terms of energy dissipation compared to their current-driven counterparts. In the supplementary material, we have shown a detailed comparison between BCB 3.0 and the latest results for both charge- and spin-based devices on two separate plots.

### 2) Performance under the Power Density Constraint

For many computing applications, power density is a critical constraint that limits the maximum operation speed of a chip. Therefore, it is desirable to investigate the throughput density under a fixed power density. Fig. 5 shows the constrained throughput of a 32-bit adder under the power density limit of 10W/cm$^2$ for a variety of devices. For low-power devices, the throughput is limited by the delay. The CMOS HP device cannot fully utilize its speed advantage due to the power density cap, and the slower but more energy-efficient HetJTFET has a better throughput.

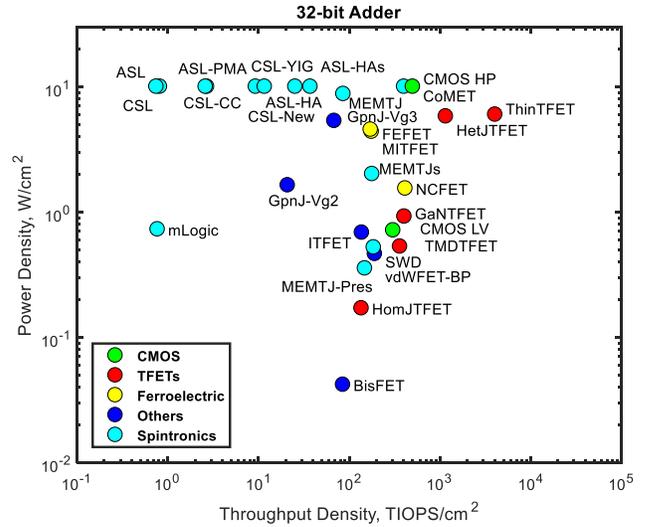

Fig. 5. Power density versus throughput density for a variety of charge- and spin-based devices.

To further improve the throughput of low-power devices, such as TFETs and voltage-controlled spintronic devices, ultra-deep pipelining circuit techniques are employed. For charge-based devices, standard N-P domino logic is implemented to enable the pipeline circuit as described in Section III A. For NDR devices and spintronic devices, supply clocking is used to achieve ultra-deep pipelining and to boost the throughput. The comparison of various technologies using deep-pipelined circuits is shown in Fig. 6. One clear trend is that low-power devices shift significantly to the top right corner. Devices closer to the power density cap benefit less from the pipelined circuit. Most charge-based devices and three voltage-controlled spintronic devices provide better throughput than the CMOS HP.

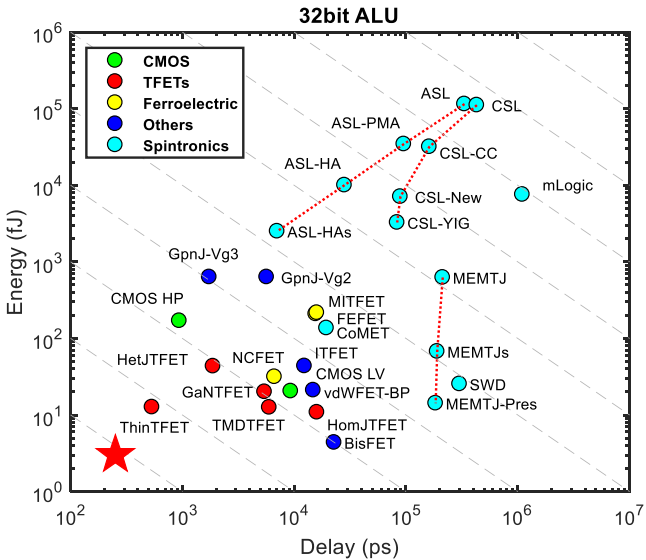

Fig. 4. Energy versus delay of a 32-bit ALU for a variety of charge- and spin-based devices. Here, ASL-HA and ASL-HAs stand for ASL devices using Heusler Alloy with nominal and improved saturation magnetization values of 4×10$^5$ and 10$^5$ A/m, respectively; CSL, CSL-CC, CSL-New, and CSL-YIG correspond to device structures shown in Fig. 1 (a) – (d), respectively; MEMTJ, MEMTJs, and MEMTJ-Preset correspond to device structures shown in Fig. 2 (a) – (c), respectively. The red star indicates the preferred corner.

For spintronic devices, the data points for the original ASL and CSL device proposals have moved away from the preferred

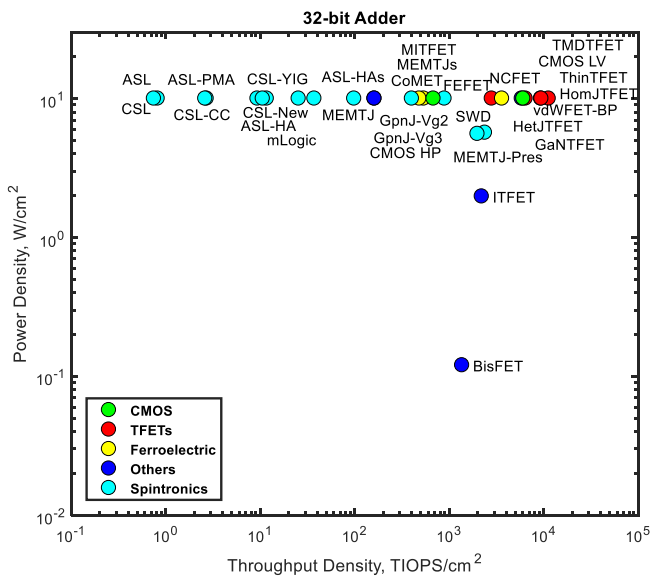

Fig. 6. Power density versus throughput density for a variety of charge- and spin-based devices with ultra-deep pipelining. The charge-based FETs are implemented with the standard N-P domino logic except for NDR devices, which inherently have a memory feature.

### B. Interconnect

#### 1) Interconnects with Repeater Insertions

Fig. 7 and Fig. 8 show the delay and energy of passing data through a 100-μm length interconnect using charge- and spin-based technologies, respectively. BisFET and ITFET are much closer to the preferred corner than other charge-based devices. This is because in the 32-bit ALU benchmarking, the leakage energy contributes to the majority of energy dissipation, as shown in Fig. 4. For the interconnect application, the dynamic energy becomes more dominant; BisFET and ITFET have the advantage of ultra-low supply voltage, leading to a much lower energy.

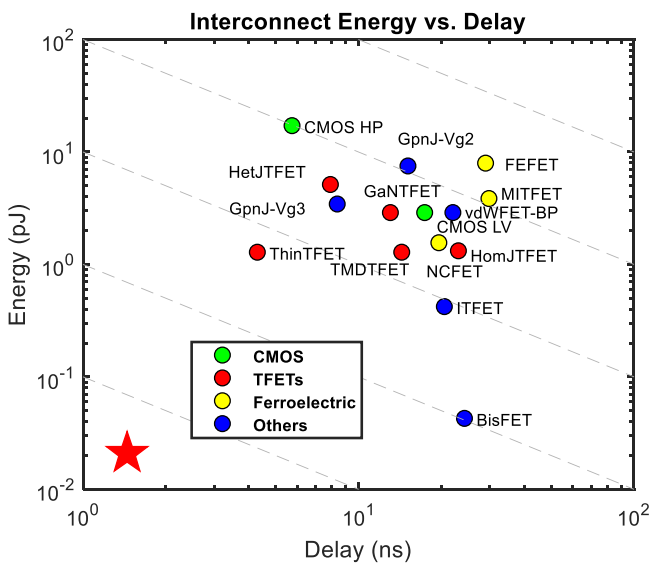

Fig. 7. Energy versus delay of a 100 um interconnect with repeater insertion using charge-based devices, where the red star indicates the preferred corner.

Due to the limited magnet switching speed, spintronic interconnects are much slower compared to charge-based interconnects, as shown in Fig. 8. Compared to results for the 32-bit ALU, the gap between the charge- and spin-based interconnects is even larger because the majority-gate-based spintronic logic is very efficient to perform a full adder. Only one majority gate is required in the critical path to generate the carry bit.

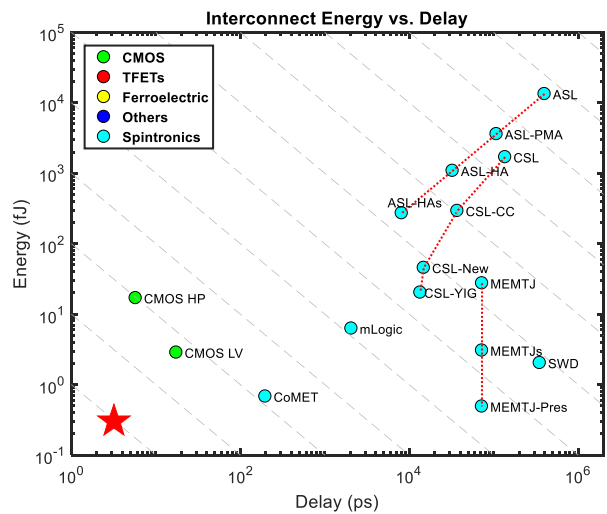

Fig. 8. Energy versus delay of a 100 um interconnect with repeater insertion using spintronic devices, where the red star indicates the preferred corner.

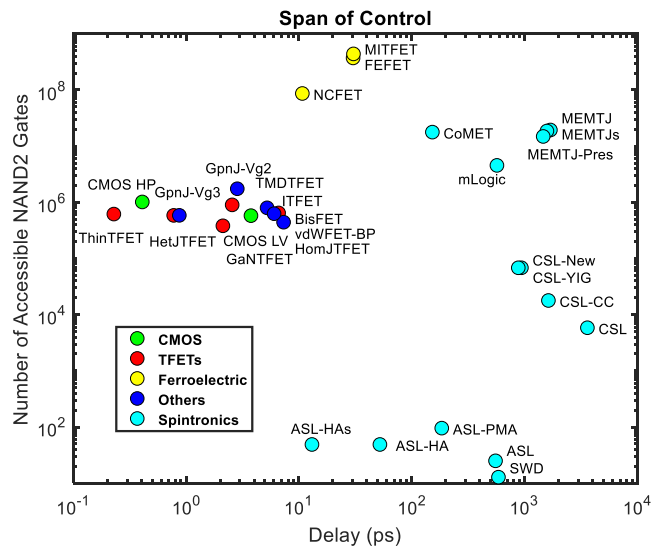

Fig. 9. Span of control versus intrinsic delay for a variety of charge- and spin-based devices.

#### 2) Span of Control

Fig. 9 shows the maximum numbers of reachable NAND2 gates per clock cycle for various emerging technologies. For instance, the GpnJ device has a relatively fast intrinsic speed and a short clock cycle; however, it has a large span and reaches more gates than most of the other devices despite a large footprint area. This is mainly because of a low resistance enabled by the high mobility of graphene. Ferroelectric- and piezoelectric-based devices, marked by yellow circles, offer the largest span of control because of their long clock cycles due to the extra polarization switching time. It should be noted that the span of control is not a performance metric; rather, it is an indicator of the circuit size beyond which interconnects impose severe limits. Obviously, an intrinsically fast device is more susceptible to performance degradation due to interconnects.

*C. Non-Boolean Computing Benchmarking based on Cellular Neural Network (CNN)*

Following our previous non-Boolean computing benchmarking for a CNN [54], the updated results are shown in Fig. 10 with the latest device-level models described in Section II. Note that to achieve 90% recall accuracy for a given input noise of 10%, the number of connectivity required is about 30, which may impose constraints on the routing. However, the benchmarking methodology in this work can be further extended to a convolutional neural network based on cellular neural network, where the connectivity requirement is 9 [13].

In Fig. 10, triangle markers show the digital CNN implementation based on CMOS HP and LV devices. Compared to analog implementation (green markers), digital CNNs require multiple addition and multiplication operations for each time step, which is time and energy consuming. Therefore, all the other emerging charge-based devices are implemented based on the analog circuits. TFETs have significant performance improvement because of their steep subthreshold slopes and large driving currents at ultra-low supply voltages. Compared to the previous benchmarking work, spintronic devices have shifted much closer to the preferred corner. This is because for spin diffusion and spin Hall effect based CNNs, a single magnet can mimic the functionality of a neuron to perform the integration; for the domain-wall based CNNs, the integration is performed by moving the domain wall inside the free magnet, which is very energy efficient due to a small critical switching current. For charge-based CNNs, an operational amplifier is required for each neuron and synapse, which consumes more power and requires a large footprint area.

## V. CONCLUSIONS

In this paper, a new release of the uniform benchmarking methodology for beyond-CMOS device is presented for both Boolean and non-Boolean logic applications. More realistic modeling approaches are included, and more advanced device material and structures are investigated. In general, spintronic devices are slower than charge-based devices because of the limited ferromagnet switching speed and domain wall propagation speed. Voltage-controlled spintronics devices are more energy-efficient than current-driven ones. Three types of cellular neural network implementations have been investigated based on a uniform benchmarking methodology. Spintronic devices show great performance in neuromorphic computing circuits, which differs significantly from their results in Boolean circuits, such as a 32-bit ALU. This indicates that new devices need to be complemented with novel circuits to achieve their full potential.

ACKNOWLEDGEMENTS

The authors would like to thank colleagues in SRC STARnet and NRI: J. Nahas, J. Appenzeller, S. Datta, C. Kim, S. Sapatnekar, M. Mankalale, Z. Liang, J. Wang, P. Dowben, S. Hu, M. Niemier, V. Narayanan, S. Salahuddin, A. Seabaugh, F. Register, A. Marshall, R. Lake, S. Sylvia, A. Ghosh, and M. Elahi. They would also like to thank the NRI/STARnet Benchmarking Steering Committee members, I. Young and D. Nikonov from Intel Co., S. Kramer from Micron, W. Haensch from IBM, and J. Herbsommer from Texas Instruments for many useful discussions during the quarterly meetings. They also acknowledge contributions from current and former colleagues at Georgia Tech, S. Chang, S. Dutta, N. Kani, R. Irai, V. Huang, and C. Hsu.

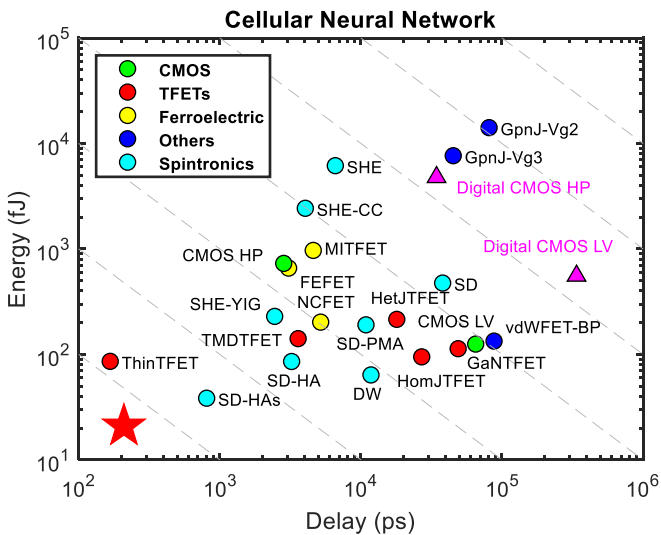

Fig. 10. Energy versus delay per memory association operation using CNN for a variety of charge- and spin-based devices, where the red star indicates the preferred corner.